\newcommand{\de}{\delta}

\newcommand{\la}{\lambda}

\newcommand{\om}{\omega}

\newcommand{\been}{\begin{equation}}
\newcommand{\een}{\end{equation}}
\newcommand{\beena}{\begin{eqnarray}}
\newcommand{\eena}{\end{eqnarray}}

\newcommand{\tn}{\textnormal}

\newcommand{\deriv}[2]{\frac{\partial{#1}}{\partial{#2}}}
\newcommand{\derivtwo}[2]{\frac{\partial^{2}{#1}}{\partial{#2}^2}}

\documentclass[aip,apl,reprint]{revtex4-1}

\usepackage{dcolumn}

\usepackage{amsmath,amssymb}
\usepackage{verbatim}
\usepackage{graphicx,color}
\usepackage{psfrag}

\DeclareMathOperator{\Div}{div}

\def\dd{\text{d}}

\begin{document}



\title[Antiplane elastic wave cloaking]{Employing pre-stress to generate finite cloaks for antiplane elastic waves}



\author{William J.\ Parnell}
\email{William.Parnell@manchester.ac.uk}
\affiliation{School of Mathematics, Alan Turing Building, University of Manchester, Oxford Road, Manchester, M13 9PL, UK}

\author{Andrew N.\ Norris}
\email{norris@rutgers.edu}
\affiliation{Mechanical and Aerospace Engineering, Rutgers University,
Piscataway, NJ 08854-8058,  USA}

\author{Tom Shearer}
\email{tom.shearer@postgrad.manchester.ac.uk }
\affiliation{School of Mathematics, Alan Turing Building, University of Manchester, Oxford Road, Manchester, M13 9PL, UK}


\date{\today}

\begin{abstract}
It is shown that nonlinear elastic pre-stress of neo-Hookean hyperelastic materials can be used as a mechanism to generate finite cloaks and thus render objects near-invisible to incoming antiplane elastic waves. This approach appears to negate the requirement for special cloaking metamaterials with inhomogeneous and anisotropic material properties in this case. These properties are induced naturally by virtue of the pre-stress. This appears to provide a mechanism for broadband cloaking since dispersive effects due to metamaterial microstructure will not arise.
\end{abstract}

\pacs{46.40.Cd, 62.20.F-, 62.30.+d}

\maketitle 


Interest in cloaking theory and its practical realization has grown significantly since the early theoretical work of Leonhardt \cite{Leo-06} and Pendry et al.\ \cite{Pen-06} in optics and electromagnetism respectively. Methods have been largely based on the idea of coordinate transformations \cite{Gre-03} which motivate the design of cloaking metamaterials that are able to guide waves around a specific region of space. Since the early work, research has focused on the possibility of cloaking in the contexts of acoustics \cite{Cum-07,Che-07,Nor-08}, surface waves in fluids \cite{Far-08}, heat transfer \cite{Gao-10}, fluid flow \cite{Urz-11} and linear elastodynamics \cite{Mil-06,Bru-09,Ami-10,Nor-11} and it is the latter application which is the concern of the present letter. In particular it was shown in \cite{Mil-06} that elastodynamic cloaking is made difficult due to the lack of invariance of Navier's equations under general coordinate transformations which retain the symmetries of the elastic modulus tensor. A special case is that of flexural waves in thin plates \cite{Far-09}. Invariance of the governing equations \textit{can} be achieved for a more specific class of transformations if assumptions are relaxed on the minor symmetries of the elastic modulus tensor as was described for the in-plane problem in \cite{Bru-09}. Cosserat materials were exploited in \cite{Nor-11}.

As  noted in \cite{Bru-09}, another special case for elastodynamics is the antiplane elastic wave problem, where cloaking can readily be achieved from a cylindrical region (using a cylindrical cloak) in two dimensions by virtue of the duality between antiplane waves and acoustics in this dimension. Consider an unbounded homogeneous elastic material with shear modulus $\mu_0$ and density $\rho_0$ and introduce a Cartesian coordinate system $(X,Y,Z)$ and cylindrical polar coordinate system $(R,\Theta,Z)$ with some common origin $\mathbf{O}$. Planar variables are related in the usual manner, $X=R\cos\Theta, Y=R\sin\Theta$. Suppose that there is a time-harmonic line source, polarized in the $Z$ direction and located at $(R_0,\Theta_0)$, with circular frequency $\omega$ and amplitude $C$ (which is a force per unit length in the $Z$ direction). This generates 
antiplane elastic waves with the only non-zero displacement component in the $Z$ direction of the form $\tn{U}=\Re[W(X,Y)\exp(-i\om t)]$. 
The displacement $W$ is governed by
\begin{align}
\nabla_{\mathbf{X}}\cdot\left(\mu_0\nabla_{\mathbf{X}} W\right) + \rho_0\om^2 W &= C\de(\mathbf{X}-\mathbf{X}_0), 
\label{antiplane}
\end{align}
where $\nabla_{\mathbf{X}}$ is the gradient operation in the ``untransformed'' frame, $\mathbf{X}=(X,Y)$ and $\mathbf{X}_0=(X_0,Y_0)$.

The assumed 
mapping for a cloak for antiplane waves (cf.\ acoustics) expressed in plane cylindrical polar coordinates, takes the form
\begin{align}
r &= g(R) , & \theta &= \Theta, & z&=Z, \quad \text{for}\ 0\leq R\leq R_2 ,
\label{1=2}
\end{align}
and the identity mapping for all $R>R_2$ for some chosen monotonically increasing function $g(R)$ with $g(0) \equiv r_1 \in [0,R_2]$, $g(R_2)=R_2\in\mathbb{R}$ such that $R_2<R_0$,
i.e.\ the line source remains outside the cloaking region. The cloaking region is thus defined by $r\in[r_1,r_2]$ where $r_2=R_2$. We use upper and lower case variables for the untransformed and transformed problems respectively. Under this mapping the form of the governing equation \eqref{antiplane} remains unchanged for $R=r>R_2$,
whereas for $0\leq R\leq R_2$, corresponding to the transformed domain $r_1\leq r\leq R_2$, the transformed equation takes the form (in transformed cylindrical polar coordinates $r,\theta=\Theta$)
\begin{align}
\frac{1}{r}\deriv{}{r}\left(r\mu_r(r)\deriv{w}{r}\right) + \frac{\mu_{\theta}(r)}{r^2}\derivtwo{w}{\theta} + d(r)\om^2 w &= 0 \label{inhomwave}
\end{align}
where (see eqs.\ (26), (27) in \cite{Nor-11})
\begin{equation}
\mu_r(r) = \frac{\mu_0^2}{\mu_\theta (r)} = \mu_0\frac Rr  \frac{\dd g}{\dd R},
\ \
d(r) = \rho_0\frac Rr \left(\frac{\dd g}{\dd R}\right)^{-1}. \label{classicalprops}
\end{equation}
Hence, \textit{both} the shear modulus and density must be inhomogeneous and the shear modulus must be anisotropic. Material properties of this form cannot be constructed exactly since the shear modulus $\mu_{\theta}$ becomes unbounded as $r\rightarrow r_1$ (the inner boundary of the cloak). In this limit the density behaves as $d= (pcr_1)^{-1}\rho_0 R^{2-p} + \ldots$ where $p,\, c >0$ define the mapping in the vicinity of the inner boundary according to $r = r_1 + c R^p + \ldots$ as $R\rightarrow 0$. In practice of course approximations are required as described in e.g.\ \cite{Sch-06,Far-08,Zha-11}. Note that, as expected \cite{Nor-11}, the total mass is conserved since, regardless of the  mapping,
the integral of the density $d(r)$ over $r\in[r_1,r_2]$ is $\pi R_2^2 \rho_0$.

In \cite{Par-11} a new method to generate elastic cloaks was proposed which used the notion of nonlinear pre-stress.
That this was possible was due to the fact that the antiplane wave field scattered from a cylindrical cavity \textit{is invariant under pre-stress for an incompressible neo-Hookean material}. Scattering coefficients in the deformed configuration depend only on the \textit{initial} cavity radius $R_1$ and therefore provided that this is small compared with the incident wavelength, scattering from the inflated cavity of radius $r_1$ will be negligible regardless of the relative size of $r_1$ and the incident wavelength. Therefore we can conclude that an object placed inside the inflated cavity region would be near-invisible (i.e.\ cloaked) upon choosing $R_1$ appropriately. In \cite{Par-11} the pre-stress affected the entire elastic domain however and therefore its influence was felt by both the source and receiver. In this letter we show how this theory may be adapted in order to create a finite cloak by means of an axial stretch.

\begin{figure}[h!]
\begin{center}
\includegraphics[scale=0.4]{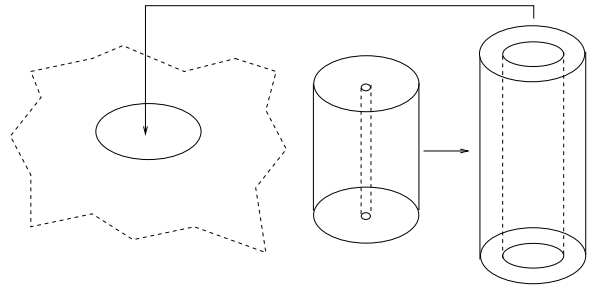}
\caption{The incompressible neo-Hookean cylindrical annulus is pre-stressed as depicted on the right. This annulus then creates a cloak when slotted into a cylindrical cavity in an unbounded elastic medium, as illustrated on the left.
}\label{figdef}
\end{center}
\end{figure}

With reference to Fig.\ \ref{figdef}, let us consider an elastic material within which is located a cylindrical cavity of radius $R_2$. Let us assume that the density of this medium is $\rho_0$ and its axial shear modulus (corresponding to shearing on planes parallel to the axis of the cylindrical cavity) is $\mu_0$.  Additionally we take a cylindrical annulus of isotropic \textit{incompressible} neo-Hookean material with associated shear modulus $\mu$ and density $\rho$ and with inner and outer radii $R_1$ and $R_2$ respectively with $R_1\ll R_2$. The exact nature of this latter relationship will be described shortly. We shall consider deformations of the cylindrical annulus in order that it can act as an elastodynamic cloak to incoming antiplane elastic waves. We deform the material so that its inner radius is significantly increased (to $r_1$) but its outer radius $R_2$ remains unchanged. The deformed cylindrical annulus can then slot into the existing cylindrical cavity region within the unbounded (unstressed) domain. We choose $\mu$ and $\rho$ so that subsequent waves satisfy the necessary continuity conditions on $r=R_2$.

The constitutive behaviour of an incompressible neo-Hookean material is described by the strain energy function \cite{Ogd-97}
\begin{align}
\mathcal{W} &=  \frac{\mu}{2}(\la_r^2+\la_{\theta}^2+\la_z^2 - 3) \label{NHSEF}
\end{align}
where $\la_j, j=r,\theta,z$ are the radial, azimuthal and axial principal stretches of the large deformation to ensue. We consider the initial deformation of the cylindrical annulus domain as depicted in Fig.\ \ref{figdef}. Since the material is incompressible, the deformation is induced \textit{either} by applying a uniform axial stretch $L$ \textit{or} a radial pressure difference $p_o-p_i$ where $p_o$ and $p_i$ denote the pressures applied to the outer and inner face of the cylindrical annulus respectively. The ensuing deformation is described via the relations
\begin{equation}
R=R(r),\qquad \Theta=\theta,\qquad Z=z/L,
\label{initial deformation}
\end{equation}
where $(R,\Theta,Z)$ and $(r,\theta,z)$ are cylindrical polar coordinates in the undeformed and deformed configurations. 
Note the convention introduced in \eqref{initial deformation}, i.e.\ that upper case variables correspond to the undeformed configuration whilst lower case corresponds to the deformed configuration. This is analogous to the notation used for untransformed and transformed configurations in \eqref{1=2}.



The principal stretches for this deformation are
\begin{equation}
\lambda_r=\frac{\dd r}{\dd R}=\frac{1}{R^\prime(r)}, \  \ \lambda_\theta=\frac{r}{R(r)},
\ \ \lambda_z=L.
\end{equation}
For an incompressible material $\la_r\la_{\theta}\la_z=1$, implying
\begin{equation}
R(r)=\sqrt{L(r^2+M)}, \label{eqn:Q(r)}
\end{equation}
where $M=R_2^2(L^{-1}-1)$ is a constant determined by imposing that the outer wall of the cylindrical annulus remains fixed, i.e. $R(R_2)=R_2$. The deformation \eqref{eqn:Q(r)} is easily inverted to obtain $r(R)$.
Given incompressibility and the fixed outer wall of the annulus, in order to induce this deformation we may either (i) prescribe the axial stretch $L$ which then determines the deformed inner radius $r_1$ and the radial pressure difference required to maintain the deformation \textit{or} (ii) prescribe the radial pressure difference which then determines the deformed inner radius $r_1$ and the axial stretch $L$.

We shall discuss the radial pressure difference shortly but either way we can obtain $L$ and thus feed this into \eqref{eqn:Q(r)}. Imposing the requirement that $R(r_1)=R_1$ and using the form of $M$ gives rise to the useful relation
\begin{align}
L &= \frac{R_2^2-R_1^2}{R_2^2-r_1^2}. \label{bigL}
\end{align}

The Cauchy stress   for an incompressible material is \cite{Ogd-97}
\begin{align}
\mathbf{T} &= \mathbf{F}\frac{\partial \mathcal{W}}{\partial\mathbf{F}}+Q\mathbf{I},
\label{Cauchy stress tensor}
\end{align}
where $\mathcal{W}$ is the neo-Hookean strain energy function introduced in \eqref{NHSEF}, $\mathbf{F}$ is the deformation gradient, $\mathbf{I}$ is the identity tensor and $Q$ is the scalar Lagrange multiplier associated with the incompressibility constraint.

Only diagonal components of the Cauchy stress are non-zero, being given by (no sum on the indices)
\begin{align}
T_{jj} &= \mu_{j}(r) + Q \label{Tjj}
\end{align}
for $j=r,\theta,z$,
where
\begin{align}
\mu_r(r) &= \frac{\mu^2}{L^2}\frac{1}{\mu_{\theta}(r)} =
\frac{\mu}{L}\left(\frac{r^2+M}{r^2}\right), & \mu_z &= L^2\mu  . \label{mu}
\end{align}
The second and third of the static equations of equilibrium $\Div\mathbf{T}=\mathbf{0}$ (where $\Div$ signifies the divergence operator in the deformed configuration) merely yield $Q=Q(r)$. The remaining equation
\begin{align}
\frac{\partial T_{rr}}{\partial r}+\frac{1}{r}(T_{rr}-T_{\theta\theta})&=0,
\end{align}
can be integrated using \eqref{Tjj}-\eqref{mu} to obtain $Q(r)$.  Writing $T_{rr}\big|_{r=R_2}=-p_{o}, T_{rr}\big|_{r=r_1}=-p_{i}$ we find
\begin{align}
\frac{L(p_{i}-p_{o})}{\mu} &= \frac{1}{2L}\left(1-\frac{R_1^2}{r_1^2}\right)
+\log\left(\frac{r_1}{R_1}\right) . \label{Trreqn2}
\end{align}
Given $L$ and thus $r_1$ via \eqref{bigL}, this equation prescribes the required pressure difference.


%


Now assume that the cylindrical annulus has been pre-stressed in an appropriate manner and slotted into the unbounded elastic material with perfect bonding at $r=R_2$. We consider wave propagation in this medium given a time-harmonic antiplane line source located at $(R_0,\Theta_0)$ with $R_0>R_2$. In $r>R_2$ the antiplane wave with corresponding displacement which we shall denote by $w(r,\theta)$, is again governed by \eqref{antiplane}. 
In the region $r_1\leq r\leq R_2$, the wave satisfies a different equation since this annulus region has been pre-stressed according to the deformation \eqref{initial deformation} and \eqref{eqn:Q(r)}.  We can obtain the governing equation using the theory of small-on-large \cite{Ogd-97}. It was shown in \cite{Par-11} that the wave in this region satisfies \eqref{inhomwave} but now where $\mu_r(r)$ and $\mu_{\theta}(r)$ are defined in \eqref{mu} and $d(r)=\rho$,
 and note that we have made the necessary corrections in order to include the axial stretch $L$ which was not considered in \cite{Par-11}. Note in particular that the density is homogeneous inside the cloak region.

Let us introduce the identity mapping for $r>R_2$ and
\begin{equation}
R^2 = L(r^2+M), \ \  \Theta=\theta,  \
\text{for}\ r_1\leq r\leq R_2
\label{mapping}
\end{equation}
which corresponds to the actual physical deformation \eqref{eqn:Q(r)}. Finally define $W(R,\Theta)=w(r(R),\theta(\Theta))$. It is then straightforward to show that the equation governing wave propagation in the entire domain $R\geq R_1$ is \eqref{antiplane}, \textit{provided that we choose $\mu=L\mu_0$ and $\rho=L\rho_0$.} These relations ensure that the wavenumbers in the exterior and cloak regions are the same and they also maintain continuity of traction on $R=R_2$.
Furthermore since \eqref{mapping} corresponds to the actual deformation, the inner radius $r_1$ maps back to $R_1$. Therefore with the appropriate choice of cloak material properties, the scattering problem in the undeformed and deformed configurations are equivalent. We can therefore solve the equation in the undeformed configuration and then map back to the deformed configuration to find the physical solution. Decomposing the solution into incident and scattered parts $W=W_i+W_s$, we have $W_i = \frac{C}{4i\mu_0}\tn{H}_0(KS)$ where we have defined the wavenumber $K$ via $K^2=\rho_0\om^2/\mu_0$ and $S=\sqrt{(X-X_0)^2+(Y-Y_0)^2}$.
 Here $\tn{H}_n=\tn{H}_n^{(1)}$ is the Hankel function of the first kind of order $n$.
The scattered field is written in the form \cite{Par-11}
\begin{align}
W_s(R) &= \sum_{n=-\infty}^{\infty}(-i)^n a_n\tn{H}_n(KR)e^{in(\Theta-\Theta_0)}. \label{WsR}
\end{align}
Satisfaction of the traction free boundary condition on $R=R_1$ gives $a_n$.
We want the wave field with respect to the \textit{deformed} configuration, so we
map back in order to find $w=w_i+w_s$. The incident wave is most conveniently determined by using Graf's addition theorem in order to distinguish between the regions $r<R_0$ and $r>R_0$, as was described in \cite{Par-11}.
The incident and scattered fields are then, respectively,
\begin{widetext}
\begin{align}
w_i(r) &=
\frac{C}{4i\mu_0}\sum_{n=-\infty}^{\infty}e^{in(\theta-\theta_0)}\times
\begin{cases}
\tn{H}_n(KR_0)J_n(K\sqrt{L(r^2+M)}), & r_1\leq r<R_2, \\
\tn{H}_n(KR_0)J_n(Kr), & R_2\leq r<R_0, \\
\tn{H}_n(Kr)J_n(KR_0), & r>R_0,
\end{cases}
\\
w_s(r) &= -\frac{C}{4i\mu_0}
\sum_{n=-\infty}^{\infty}  e^{in(\theta-\theta_0)}
\, \frac{\tn{J}_n'(KR_1)}{\tn{H}_n'(KR_1)}\tn{H}_n(KR_0)
\times \begin{cases}
\tn{H}_n\left(K\sqrt{L(r^2+M)}\right), & r_1\leq r<R_2, \\
\tn{H}_n\left(Kr\right), \label{wssol} & r\geq R_2.
\end{cases}
\end{align}

\end{widetext}

\vspace{-0.5cm}

\begin{figure}[h!]
\begin{center}
\includegraphics[scale=0.82]{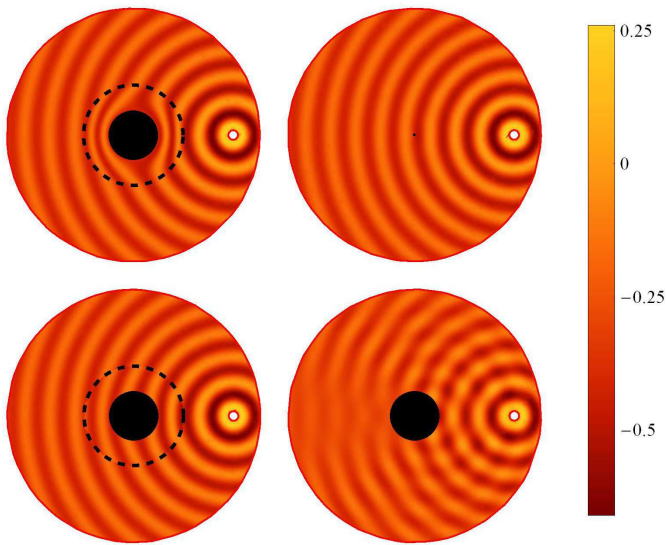}
\caption{Cloaking of antiplane shear waves. Line source is located at $Kr=KR_0=8\pi$, $\Theta_0=0$, shown as a white circle. \textbf{Upper left}: A region of (nondimensionalized) radius $Kr_1=2\pi$ is cloaked using a classic linear elastic cloak $g(R) = r_1 +R\big(\frac{R_2-r_1}{R_2}\big)$ in $2\pi\leq Kr\leq 4\pi$. \textbf{Upper right}: Scattering from a cavity of radius $KR_1=2\pi/20$ in an unstressed medium. \textbf{Lower left}: A ``pre-stress'' cloak in $2\pi\leq Kr\leq 4\pi$ generated from an annulus with initial inner radius $KR_1=2\pi/20$. \textbf{Lower right}: Scattering from a cavity with radius $KR_1=2\pi$ in an unstressed medium. Scattering and the shadow region presence in the latter is significant, as compared with that for an equivalent sized cavity for the ``pre-stress'' cloak.}
\label{fig2}
\end{center}
\end{figure}

\vspace{-0.3cm}

The key to cloaking is to ensure that the scattered field is small compared with the incident field, i.e.\ $a_n\ll 1$. Note from \eqref{wssol} that $a_n$ are solely dependent on the initial annulus inner radius $R_1$ (and source distance $R_0$) but are \textit{independent} of the deformed inner radius $r_1$. Therefore we must choose $R_1$ such that $KR_1\ll 1$ which will ensure negligible scattering. We illustrate with some examples in Fig.\ \ref{fig2}, showing that the ``pre-stress'' cloak appears to work well.

In conclusion, we have shown how a finite cloak for antiplane elastic waves can be generated by employing nonlinear pre-stress of an incompressible neo-Hookean hyperelastic material. The performance of the cloak is limited only by the size of the initial radius of the cylindrical cavity inside the annulus region. The anisotropic, inhomogeneous material moduli in the cloaking region, defined by \eqref{mu}, are induced naturally by the pre-stress and therefore exotic metamaterials are not required. Dispersive effects, which naturally arise in metamaterials due to their inherent inhomogeneity at some length scale, will not be present in the pre-stress context and we also note that the density of the cloak is homogeneous. In order to achieve the required pre-stress, a radial pressure difference is required across the cylindrical annulus. It would be inconvenient to prescribe $p_o$ on the outer face. However, since we only need a pressure \textit{difference} we can prescribe $p_i$ with $p_o=0$, ensuring the prescribed deformation and eliminating this difficulty. The incompressible neo-Hookean model is an approximation to reality, holding in general for rubber-like materials and moderate deformations. If the material is \textit{not} neo-Hookean, invariance of the scattering coefficients is not guaranteed in general and therefore similar exact results will not hold. However it would be of interest to ascertain whether scattering from inflated cavities in other hyperelastic pre-stressed media is still significantly reduced as compared with an equivalent sized cavity in an unstressed medium.

In closing we remark that one of the fundamental advantages of the pre-stress approach is that pre-stress generates equations with incremental moduli (analogies of the elastic moduli) which do \textit{not} possess the minor symmetries. Therefore this approach can be used for cloaking in the more general elastodynamic setting where classical linear elastic materials cannot be used\cite{Nor-12}.



%
%

%

\vspace{-0.6cm}

\begin{acknowledgments}

\vspace{-0.3cm}

The work of ANN was supported by NSF and ONR. TS is grateful to EPSRC for funding his PhD research studentship.
\end{acknowledgments}

\vspace{-0.2cm}



\end{document}